\documentstyle[epsf]{aipproc}

\def\ra{\rightarrow}

\catcode`@=11
\long\def\@makefntext#1{\parindent 0pt\hsize\columnwidth\parskip0pt\relax
\footnotesize\baselineskip12pt\def\strut{\vrule width0pt height0pt depth1.75pt\relax}%
\mbox{$\m@th^{\@thefnmark}$\hspace*{3pt}}#1}
\catcode`@=12

\begin{document}
\pagestyle{plain}

\font\fortssbx=cmssbx10 scaled \magstep2
\hbox to \hsize{{  }  
\hfill
\vtop{\hbox{\bf UH-511-966-00}
     \hbox{\bf August 2000}}
}

\title{Testing Neutrino Properties at Long Baseline Experiments and
Neutrino Factories\footnote{Presented at the NuFACT'00, Monterey, May
22-26, 2000, to be published in the proceedings.}}

\author{Sandip Pakvasa}

\address{\vglue-3ex
Department of Physics \& Astronomy  \\
University of Hawaii  \\ 
Honolulu, HI 96822\vglue-3ex}

\maketitle

%

%
%

\begin{abstract}
It is shown that explanations of atmospheric neutrino anomaly other than
$\nu_\mu-\nu_\tau$
oscillations (e.g. decay, decoherence and $\nu_\mu-\nu_\tau- \nu_{KK}$ 
mixing) can be tested
at future facilities.  Stringent tests of CPT invariance in neutrino
oscillations can also be performed.  
\end{abstract}

\section{Introduction}

In this talk I would like to (a) review some non-oscillatory
explanations for the atmospheric neutrino data, and how they can be
distinguished from the conventional oscillatory explanations in future
neutrino experiments; and (b) describe briefly the strong limits that can
be placed on CPT violation both from existing data as well as future
experiments especially at neutrino factories.

\section{Neutrino Decay, Decoherence and Extra Dimensions}

Decay

If neutrinos do have masses and mixings; then in general, in addition
to oscillating, the heavier neutrinos will decay to lighter ones via
flavor changing processes.  The only questions are (a) whether the
lifetimes are short enough to be interesting and (b) what are the
dominant decay modes.  To be specific, let us assume that neutrino masses
are at most of order eV[1].

For eV neutrinos, the only radiative mode possible is       
$\nu_i \ra \nu_j + \gamma$.
 From the existing bounds on neutrino magnetic moments, indirect (but
model independent) bounds on the decay rates for this mode can be
derived: $10^{-11} s^{-1}, 10^{-17} s^{-1}$ and $10^{-19} s^{-1}$
for $\nu_\tau, \nu_\mu$ and $\nu_e$ respectively[1].

The decay rate for the invisible three body decay $\nu_i \ra \nu_j \nu
\bar{\nu}$ can be written (for $m_i \gg m_j)$ as
\begin{equation}
\Gamma  =  \frac{\epsilon^2_i G^2_F \ m_i^5}
                {192 \pi^3}
\end{equation}
The current experimental bound on $\epsilon_\mu$ is of 0(100)[2] (the one
loop result in SM is $\epsilon^2 \stackrel{\sim}{<} 3.10^{-12})$ and
thus $\Gamma$ for $m_i \sim 0(eV)$ has to be less than $10^{-30}$
s$^{-1}$.

The possible existence of a $I_w=0$, $J=0$ and $L=0$ massless particle, $\chi$
(such as a Nambu-Goldstone boson of broken family symmetry) leads to a
flavor changing decay mode:
\begin{equation}
\nu{_{\alpha_{L}}} \ra \nu{_{\beta_{L}}}  +  \chi
\end{equation}
By $SU(2)_L$ symmetry decays $\ell_\alpha \ra \ell_\beta + \chi$ will occur
with the same strength.  Current bounds on $BR(\mu \ra e \chi)$ and on
$BR (\tau \ra \mu \chi)$ of $2.10^{-6}$ and 
$7.10^{-6}$[3] respectively are sufficient to constrain 
$\nu_\mu$ and $\nu_\tau$ lifetimes to be longer than $10^{29}$ s
and $10^{20}$ s.

The only possibility for fast invisible decays of neutrinos seems to lie
with majoron models.  There are two classes of models; the I=1
Gelmini-Roncadelli
[4] majoron and the 
I=0 Chikasige-Mohapatra-Peccei[5] majoron.  In general, one can choose the majoron to be a mixture of the
two; furthermore the coupling can be to flavor as well as sterile
neutrinos.  The effective interaction is of the form:
\begin{equation}
g_\alpha \bar{\nu}_\beta^c \nu_\alpha \ J
\end{equation}
giving rise to decay:
\begin{equation}
\nu_\alpha \ra \bar{\nu}_\beta \ + J
\end{equation}
where $J$ is a massless $J=0 \ L=2$ particle; $\nu_\alpha$ and $\nu_\beta$
are mass eigenstates which may be mixtures of flavor and sterile
neutrinos.  Models of this kind which can give rise to fast neutrino
decays and satisfy the bounds below have been discussed by Valle,
Joshipura and others[6]. 
These models are unconstrained by $\mu$ and
$\tau$ decays which do not arise due to the $\Delta L=2$ nature of the
coupling.  The I=I coupling is constrained by the bound on the invisible
$Z$ width;
and requires that the Majoron be a mixture of I=1 and I=0.
The couplings of $\nu_\mu$ and $\nu_e \ (g_\mu$ and $g_e)$ are
constrained by the limits on multi-body $\pi, K$ decays ${\pi \ra \mu
\nu \nu  \nu}$ and $K \ra \mu \nu \nu \nu$ and on $\mu-e$ university
violation in $\pi$ and K decays[7].

Granting that models with fast, invisible decays of neutrinos can be
constructed, can such decay modes be responsible for any observed neutrino anomaly?

 We assume a
component of $\nu_\alpha,$ i.e., $\nu_2$, to be the only unstable state,
with a rest-frame lifetime $\tau_0$, and we assume two flavor mixing,
for simplicity:
\begin{equation}
\nu_\alpha = cos \theta \nu_2 \ + sin \theta \nu_1
\end{equation}
with $m_2 > m_1$.  From Eq. (2) with an unstable $\nu_2$, the $\nu_\alpha$
survival probability is
\begin{eqnarray}
P_{\alpha \alpha} &=& sin^4 \theta \ + cos^4 \theta {\rm exp} (-\alpha L/E)
            \\ \nonumber
&+& 2 sin^2 \theta cos^2 \theta {\rm exp} (-\alpha L/2E)
            cos (\delta m^2 L/2E),
\end{eqnarray}
where $\delta m^2 = m^2_2 - m_1^2$ and $\alpha = m_2/ \tau_0$.
Since we are attempting to explain neutrino data without oscillations
there are two appropriate limits of interest.  One is when the $\delta
m^2$ is so large that the cosine term averages to 0.  Then the survival
probability becomes
\begin{equation}
P_{\mu\mu} = sin^4 \theta \ + cos^4 \theta {\rm exp} (-\alpha L/E)
\end{equation}
Let this be called decay scenario A.  The other possibility is when
$\delta m^2$ is so small that the cosine term is 1, leading to a
survival probability of 
\begin{equation}
P_{\mu \mu} = (sin^2 \theta + cos^2 \theta {\rm exp} (-\alpha L/2E))^2
\end{equation}

We note in passing that scenario A does not provide an acceptable fit to
atmospheric neutrino data [8,9]. 
Turning to decay scenario B, consider the following possibility[10].
The three states $\nu_\mu, \nu_\tau, \nu_s$ (where $\nu_s$ is a
sterile neutrino) are related to the mass eigenstates $\nu_2, \nu_3,
\nu_4$ by the approximate mixing matrix
\begin{equation}
\left( \begin{array}{c} \nu_\mu\\ \nu_\tau\\ \nu_s \end{array} \right) =
\left( \begin{array}{ccc}  \cos\theta& \sin\theta& 0\\
                          -\sin\theta& \cos\theta& 0\\
                           0& 0& 1
\end{array} \right)
\left( \begin{array}{c} \nu_2\\ \nu_3\\ \nu_4 \end{array} \right)
\label{eq:mixing}
\end{equation}
and the decay is $\nu_2 \to \bar\nu_4 + J$. The electron neutrino,
which we identify with $\nu_1$, cannot mix very much with the other
three because of the more stringent bounds on its couplings\cite{barger1},
and thus our preferred solution for solar neutrinos would be
small angle matter oscillations.

In this case the $\delta m_{23}^2$ in Eq.~(6) is not
related to the $\delta m_{24}^2$ in the decay, and can be very small,
say $ < 10^{-4}\rm\, eV^2$ (to ensure that oscillations play no role in the atmospheric
neutrinos). Then the oscillating term is 1 and $P(\nu_\mu\to
\nu_\mu)$ is given by Eq. (8).

The decay  model of Equation (8) above gives a very good fit to the
Super-K data [11]
with a minimum $\chi^2 = 33.7$ (32 d.o.f.)
for the choice  of  parameters
\begin{equation}
\tau_\nu/m_\nu = 63\rm~km/GeV,
\ \cos^2 \theta = 0.30
\end{equation}
and normalization $\beta = 1.17$.

The fits (as shown in Fig. 1 in Ref. 10) show the
ratios between the Super-K data and the Monte Carlo  predictions
calculated  in the  absence of oscillations or other
form of `new physics' beyond the standard model.
The  best fits  of the two  models
 (viz. $\nu_\mu -\nu_\tau$ oscillations and decay) are  
of comparable  quality.
The reason  for the similarity  of the results  obtained
in  the two  models  can be understood by looking  at
Fig.~1, where  I show
the survival probability $P(\nu_\mu \to \nu_\mu)$
of muon neutrinos   as  a  function
of $L/E_\nu$ for  the  two  models   using the
best  fit  parameters.
In the case  of the neutrino  decay model   (thick  curve)
the probability   $P(\nu_\mu \to \nu_\mu)$
monotonically  decreases   from    unity  to  an  asymptotic  value
$\sin^4 \theta \simeq  0.49$.
In the case of  oscillations the  probability  has  a sinusoidal
behaviour  in $L/E_\nu$.  The  two  functional    forms
seem     very different;  however,  taking  into  account  the
resolution in $L/E_\nu$,  the  two  forms
are  hardly  distinguishable.
In fact, in the    large $L/E_\nu$    region, the oscillations
are  averaged  out  and the survival  probability there
can  be  well  approximated  with 0.5  (for  maximal  mixing).
In  the region  of  small  $L/E_\nu$  both probabilities  approach
unity.
In the region $L/E_\nu$ around  400~km/GeV, where  the  probability for the
neutrino oscillation model  has the first  minimum,
the  two  curves are  most  easily  distinguishable, at least in
principle.

For the atmospheric neutrinos in Super-K, two kinds of tests have been proposed
to distinguish between $\nu_\mu$--$\nu_\tau$ oscillations and
$\nu_\mu$--$\nu_s$ oscillations. One is based on the fact that matter effects
are present for $\nu_\mu$--$\nu_s$ oscillations [12] 
but are nearly absent for
$\nu_\mu$--$\nu_\tau$ oscillations [13] 
leading to differences in the zenith angle 
distributions  due to
matter effects on upgoing neutrinos [14]. 
In our case since the mixing is $\nu_\mu - \nu_\tau$  
no matter effect is expected; and hence the recent Super-K results[15] are
in accord with expectations of this decay model.  The other test is
based on the neutral current rate (as measured via   production or
multi-ring events) which is unaffected in $\nu_\mu-\nu_\tau$ 
oscillations but reduced in $\nu_\mu -\nu_s$
oscillations[16].  In our case of the decay model, the neutral current rate
is affected and the expectation is closer to $\nu_\mu-\nu_s$ mixing.

\medskip
\noindent{Long-Baseline Experiments}

The survival probability of $\nu_\mu$ as a function of $L/E$ is given in
Eq.~(1). The conversion probability into $\nu_\tau$ is given by
\begin{equation}
P(\nu_\mu\to\nu_\tau) = \sin^2\theta \cos^2\theta (1-e^{-\alpha L/2E})^2 \,.
\end{equation}
This result differs from $1-P(\nu_\mu\to\nu_\mu)$ and hence is different from
$\nu_\mu$--$\nu_\tau$ oscillations. Furthermore, $P(\nu_\mu\to\nu_\mu)
+ P (\nu_\mu\to \nu_\tau)$ is
not 1 but is given by
\begin{equation}
P (\nu_\mu\to\nu_\mu) + P(\nu_\mu\to\nu_\tau) = 1 - \cos^2\theta (1 -
e^{-\alpha L/E})
\end{equation}
and determines the amount by which the predicted neutral-current rates are
affected compared to the no oscillations (or the $\nu_\mu$--$\nu_\tau$
oscillations) case.
Fig.~2 shows the results for $P(\nu_\mu\to\nu_\mu)$,
$P(\nu_\mu\to\nu_\tau)$ and $P(\nu_\mu\to\nu_\mu) +
P(\nu_\mu\to\nu_\tau)$ for the decay model and compare them to the
$\nu_\mu$--$\nu_\tau$ oscillations, for both the K2K[16]
and  MINOS[17]
(or the corresponding European project[18])
long-baseline experiments, with the oscillation and decay parameters as
determined in the fits above.

The K2K experiment, already underway, has a low energy beam $E_\nu
\approx 1\mbox{--}2$~GeV and a baseline $L=250$~km.  The MINOS experiment will have
3 different beams, with average energies $E_\nu = 3,$ 6 and 12 GeV and a
baseline $L=732$~km.  The approximate $L/E_\nu$ ranges are thus 125--250~km/GeV for
K2K and 50--250~km/GeV for MINOS.  The comparisons in Figure 2 show that the
energy dependence of $\nu_\mu$ survival probability and the neutral
current rate can both distinguish between the decay and the oscillation
models.  ICANOE and especially MONOLITH can also  test for the
oscillation dip[19].

Decoherence[20]

There are several different possibilities that can give rise to decoherence
of the neutrino beam.  An obvious one is violation of quantum mechanics,
others are unknown (flavor specific) new interactions with environment
etc[21].
Quantum gravity effects are also expected to lead to effective
decoherence[22,23].

The density matrix describing the neutrinos no longer satisfies the
usual equation of motion:
\begin{equation}
\dot{\rho}  =  -i [H, \rho]
\end{equation}
but rather is modified to
\begin{equation}
\dot{\rho} = -i [H,\rho] + D (\rho)
\end{equation}

Imposing reasonable conditions on $D(\rho)$[24] it was shown by Lisi et
al.[20] that the
$\nu_\mu$ survival probability  $P_{\mu\mu}$ has the form:
\begin{equation}
P_{\mu \mu} = cos^2 2 \theta + sin^2 2 \theta \ e^{- \gamma L} cos
\left (
\frac{\delta m^2 L}{2 E} \right ).
\end{equation}
where $\gamma$ is the decoherence parameter.  If $\delta m^2$ is very small
$(\delta m^2L/2E \ll 1)$, this reduces to
\begin{equation}
P_{\mu\mu} = cos^2 2 \theta  \ + sin^{2} \ 2 \theta \ e^{-\gamma L}
\end{equation}
If $\gamma = \alpha/E$ with $\alpha$ constant, 
then an excellent fit to the Super-K data can be
obtained with $\theta = \pi/4$ and 
$\alpha \sim 7.10^{-3}$ GeV/Km.  
(If gamma is a constant, no fit is possible and
gamma can be bounded by $10^{-22}$ GeV).  The fits to Super-K data are shown
in ref. 20. and they are as good as the decay or 
$\nu_\mu - \nu_\tau$ oscillations[25].  
The shape of
$P_{\mu\mu}$ as a function of L/E is very similar to the decay case as shown in
Fig. 1.

Large Extra Dimensions 

Recently the possibility that SM singlets propagate in extra dimensions
with relatively large radii has received some attention[26].  In addition to
the graviton, right handed neutrino is an obvious candidate to propagate
in some extra dimensions.  The smallness of neutrino mass (for a Dirac
neutrino) can be linked to this property of the right handed singlet
neutrino[27]. The implications for neutrino masses and oscillations in
various scenarios have been discussed extensively [28,29].  I focus on one
particularly interesting possibility for atmospheric neutrinos raised by
Barbieri et al [30].  The survival probability $P_{\mu \mu}$ is given by
\begin{equation}
P_{\mu\mu} (L) = \mid \Sigma^3_{\i=1}   V_{\alpha i}
V_{\alpha i}^*  A_i (L) \ \mid^2.
\end{equation}
where
\begin{equation}
A_i (L) = \Sigma^\infty_{n=0} U_{on}{^{{(i)}^2}} \ exp(i\lambda_ n{^{{(i)}^2}}L/2ER^2)
\end{equation}
where $n$ runs over the tower of Kaluza-Klein states, 
$\lambda^{(i)}{_{n}}/R^2$  are the
eigenvalues of the mass-squared matrix and $U_{on}^{(i)} (\approx 
1/\pi^2 \xi^2)$ are the matrix 
elements of the diagonalizing unitary matrix.

An excellent fit to the atmospheric neutrino data can be obtained with
the following choice of parameters:
\begin{equation}
\xi_3 = m_3 R \sim 3,    1/R \sim  10^{-3} eV,  V_{\mu 3}^2 \approx
0.4.
\end{equation}
The fit to Super-K data is shown in Ref.27  and obviously it is as good as
oscillations.  This case corresponds to $\nu_\mu$  oscillating into
$\nu_\tau$ and a large
number(about 25) of Kaluza-Klein states.  Because of the mixing with a large
number of closely spaced states, the dip in oscillations gets washed out
and $P_{\mu\mu}$ looks very much like the decay model as shown in the
Fig. 1. 

\section{CPT Violation in Neutrino~Oscillations~[31]}

Consequences of $CP$, $T$ and $CPT$ violation for neutrino oscillations
have been written down before\cite{K}. We summarize them briefly for the
$\nu_\alpha\to\nu_\beta$ flavor oscillation probabilities
$P_{\alpha\beta}$ at a distance $L$ from the source. If
\begin{equation}
P_{\alpha\beta}(L) \neq P_{\bar\alpha\bar\beta}(L) \,,
\qquad \beta \ne \alpha \,,
\end{equation}
then $CP$ is not conserved.
If
\begin{equation}
P_{\alpha\beta}(L) \neq P_{\beta\alpha}(L) \,,
\qquad \beta \ne \alpha \,,
\end{equation}
then $T$-invariance is violated.
If 
\begin{eqnarray}
P_{\alpha\beta}(L) &\neq& P_{\bar\beta\bar\alpha}(L)\,,
\qquad \beta \ne \alpha \,,
\\
\noalign{\hbox{or}}
P_{\alpha\alpha}(L) &\neq& P_{\bar\alpha\bar\alpha}(L) \,,
\end{eqnarray}
then $CPT$ is violated.
When neutrinos propagate in matter, matter effects give rise to apparent  
$CP$ {\it and} $CPT$ violation even if the mass matrix is $CP$ conserving.

The $CPT$ violating terms can be Lorentz-invariance violating (LV) or
Lorentz invariant. The Lorentz-invariance violating,
$CPT$ violating case has been discussed by Colladay and Kostelecky\cite{B}
and by Coleman and Glashow\cite{A}.

The effective LV $CPT$ violating interaction for neutrinos is of the form
\begin{equation}
\bar\nu_L^\alpha b_\mu^{\alpha\beta} \gamma_\mu \nu_L^\beta \,,
\label{eq:LV}
\end{equation}
where $\alpha$ and $\beta$ are flavor indices. We assume rotational
invariance in the ``preferred'' frame, in which the cosmic microwave
background radiation is isotropic (following Coleman and
Glashow~\cite{A}).
\begin{equation}
m^2/2p + b_0 \,,
\end{equation}
where $b_0$ is a hermitian matrix, hereafter labeled $b$.

In the two-flavor case the neutrino phases may be chosen such that
$b$ is real, in which case the interaction in Eq.~(\ref{eq:LV}) is
$CPT$ odd. The survival probabilities for flavors $\alpha$ and
$\bar\alpha$ produced at $t=0$ are given by\cite{A}
\begin{eqnarray}
P_{\alpha\alpha}(L) &=&
1 - \sin^22\Theta \sin^2(\Delta L/4)\,, \label{eq:foo}\\
\noalign{\hbox{and}}
P_{\bar\alpha\bar\alpha}(L) &=&
1 - \sin^2 2\bar\Theta \sin^2(\bar\Delta L/4) \,,\\ 
\noalign{\hbox{where}}
\Delta\sin2\Theta &=&
\left| (\delta m^2/E) \sin2\theta_m
+ 2\delta b e^{i\eta} \sin2\theta_b \right| \,,
\label{eq:delsin}\\
\Delta\cos2\Theta &=&
(\delta m^2/E) \cos2\theta_m + 2\delta b \cos2\theta_b \,.
\label{eq:delcos}
\end{eqnarray}
$\bar\Delta$ and $\bar\Theta$ are defined by similar equations with
$\delta b\to -\delta b$.  Here $\theta_m$ and $\theta_b$ define the
rotation angles that diagonalize $m^2$ and $b$, respectively, $\delta
m^2 = m_2^2 - m_1^2$ and $\delta b = b_2 - b_1$, where $m_i^2$ and $b_i$
are the respective eigenvalues. We use the convention that
$\cos2\theta_m$ and $\cos2\theta_b$ are positive and that $\delta m^2$
and $\delta b$ can have either sign.  The phase $\eta$ in
Eq.~(\ref{eq:delsin}) is the difference of the phases in the unitary
matrices that diagonalize $\delta m^2$ and $\delta b$; only one of these
two phases can be absorbed by a redefinition of the neutrino states.

Observable $CPT$-violation in the two-flavor case is a consequence of
the interference of the $\delta m^2$ terms (which are $CPT$-even) and
the LV terms in Eq.~(\ref{eq:LV}) (which are $CPT$-odd); if $\delta m^2
= 0$ or $\delta b = 0$, then there is no observable $CPT$-violating
effect in neutrino oscillations.
If $\delta m^2/E \gg 2\delta b$ then
$\Theta \simeq \theta_m$ and $\Delta \simeq \delta m^2/E$, whereas if
$\delta m^2/E \ll 2\delta b$ then $\Theta \simeq \theta_b$ and $\Delta
\simeq 2\delta b$. Hence the effective mixing angle and oscillation
wavelength can vary dramatically with $E$ for appropriate values of
$\delta b$.

We note that a $CPT$-odd resonance for neutrinos ($\sin^22\Theta = 1$)
occurs whenever $\cos2\Theta = 0$ or
\begin{equation}
(\delta m^2/E) \cos2\theta_m + 2\delta b \cos2\theta_b = 0\,;
\end{equation}
similar to the resonance due to matter effects~\cite{F,G}. The condition
for antineutrinos is the same except $\delta b$ is replaced by $-\delta
b$. The resonance occurs for neutrinos if $\delta m^2$ and $\delta b$
have the opposite sign, and for antineutrinos if they have the same
sign. A resonance can occur even when $\theta_m$ and $\theta_b$ are both
small, and for all values of $\eta$; if $\theta_m = \theta_b$, a
resonance can occur only if $\eta \ne 0$. 
If one of $\nu_\alpha$ or $\nu_\beta$ is $\nu_e$, then matter effects
have to be included. 
%

If $\eta=0$, then
\begin{eqnarray}
\Theta &=& \theta \,,
\label{eq:tan}\\
\Delta &=& (\delta m^2/E) + 2\delta b \,.
\label{eq:delta}
\end{eqnarray}
In this case a resonance is not
possible. The oscillation probabilities become
\begin{eqnarray}
P_{\alpha\alpha}(L) &=& 1 - \sin^2 2\theta \sin^2 \left\{ \left( {\delta m^2  
\over 4E} + {\delta b\over 2} \right) L \right\} \,,
\label{eq:P}\\
P_{\bar\alpha\bar\alpha}(L) &=& 1 - \sin^2 2\theta \sin^2 \left\{ \left(  
{\delta m^2 \over 4E} - {\delta b\over 2} \right) L \right\} \,.
\label{eq:Pbar}
\end{eqnarray}
For fixed $E$, the $\delta b$ terms act as a phase shift in the
oscillation argument; for fixed $L$, the $\delta b$ terms act as a
modification of the oscillation wavelength.

An approximate direct limit on $\delta b$ when $\alpha = \mu$ can be
obtained by noting that in atmospheric neutrino data the flux of
downward going $\nu_\mu$ is not  depleted whereas that of upward going
$\nu_\mu$~is[11]. Hence, the oscillation arguments in
Eqs.~(\ref{eq:P}) and (\ref{eq:Pbar}) cannot have fully developed for
downward neutrinos. Taking $|\delta b L/2| < \pi/2$ with $L\sim20$~km
for downward events leads to the upper bound $|\delta b| <
3\times10^{-20}$~GeV; upward going events could in principle test
$|\delta b|$ as low as $5\times10^{-23}$~GeV.  Since the $CPT$-odd
oscillation argument depends on $L$ and the ordinary oscillation
argument on $L/E$, improved direct limits could be obtained by a
dedicated study of the energy and zenith angle dependence of the
atmospheric neutrino data.

The difference between $P_{\alpha\alpha}$ and $P_{\bar\alpha\bar\alpha}$
\begin{equation}
P_{\alpha\alpha}(L) - P_{\bar\alpha\bar\alpha}(L) =
- 2 \sin^22\theta \sin\left({\delta m^2 L\over2E}\right) \sin(\delta b L) 
\,, \label{eq:deltaP}
\end{equation}
can be used to test for $CPT$-violation. In a neutrino factory, the
ratio of $\bar\nu_\mu \to \bar\nu_\mu$ to $\nu_\mu \to \nu_\mu$ events
will differ from the standard model (or any local quantum field theory
model) value if $CPT$ is violated. Fig. 3 shows the
event ratios $N(\bar\nu_\mu \to \bar\nu_\mu)/N(\nu_\mu \to \nu_\mu)$
versus $\delta b$ for a neutrino factory with 10$^{19}$ stored muons and
a 10~kt detector at several values of stored muon energy, assuming
$\delta m^2 = 3.5\times10^{-3}$~eV$^2$ and $\sin^22\theta = 1.0$, as
indicated by the atmospheric neutrino data[11]. The error bars
in Fig. 3  are representative statistical
uncertainties. The node near $\delta b = 8\times10^{-22}$~GeV is a
consequence of the fact that $P_{\alpha\alpha} =
P_{\bar\alpha\bar\alpha}$, independent of $E$, whenever $\delta b L = n
\pi$, where $n$ is any integer; the node in Fig.~3 is for
$n=1$. A $3\sigma$ $CPT$ violation effect is possible in such an
experiment for $\delta b$ as low as $3\times10^{-23}$~GeV for stored
muon energies of 20~GeV. Although matter effects also induce an apparent
$CPT$-violating effect, the dominant oscillation here is $\nu_\mu \to
\nu_\tau$, which has no matter corrections in the two-neutrino
limit; in any event, the matter effect is in general small for distances
much shorter than the Earth's radius.

We have also checked the observability of $CPT$ violation at other
distances, assuming the same neutrino factory parameters used above.
For $L=250$~km, the $\delta b L$ oscillation argument in
Eq.~(\ref{eq:deltaP}) has not fully developed and the ratio of $\bar\nu$
to $\nu$ events is still relatively close to the standard model value.
For $L=2900$~km, a $\delta b$ as low as $10^{-23}$~GeV may be observable
at the $3\sigma$ level. However, longer distances may
also have matter effects that simulate $CPT$ violation.

\section{Summary}

At Long Baseline Experiments and Neutrino Factories true signatures of
oscillations (dips) can be established and decay like scenarios can be
excluded with confidence.  Furthermore these facilities can test CPT
conservation at levels better than $10^{23}$ GeV.

Acknowledgments

I thank Vernon Barger, John Learned, Eligio Lisi, Paolo Lipari,
Maurizio Lusignoli, Tom Weiler and Kerry Whisnant for extensive 
discussions and collaboration. This work was supported in part 
by U.S.D.O.E under grant DE-FG 03-94ER40833.

\begin{figure}[htp]
\centerline{\epsfysize=5.5truein\epsfbox{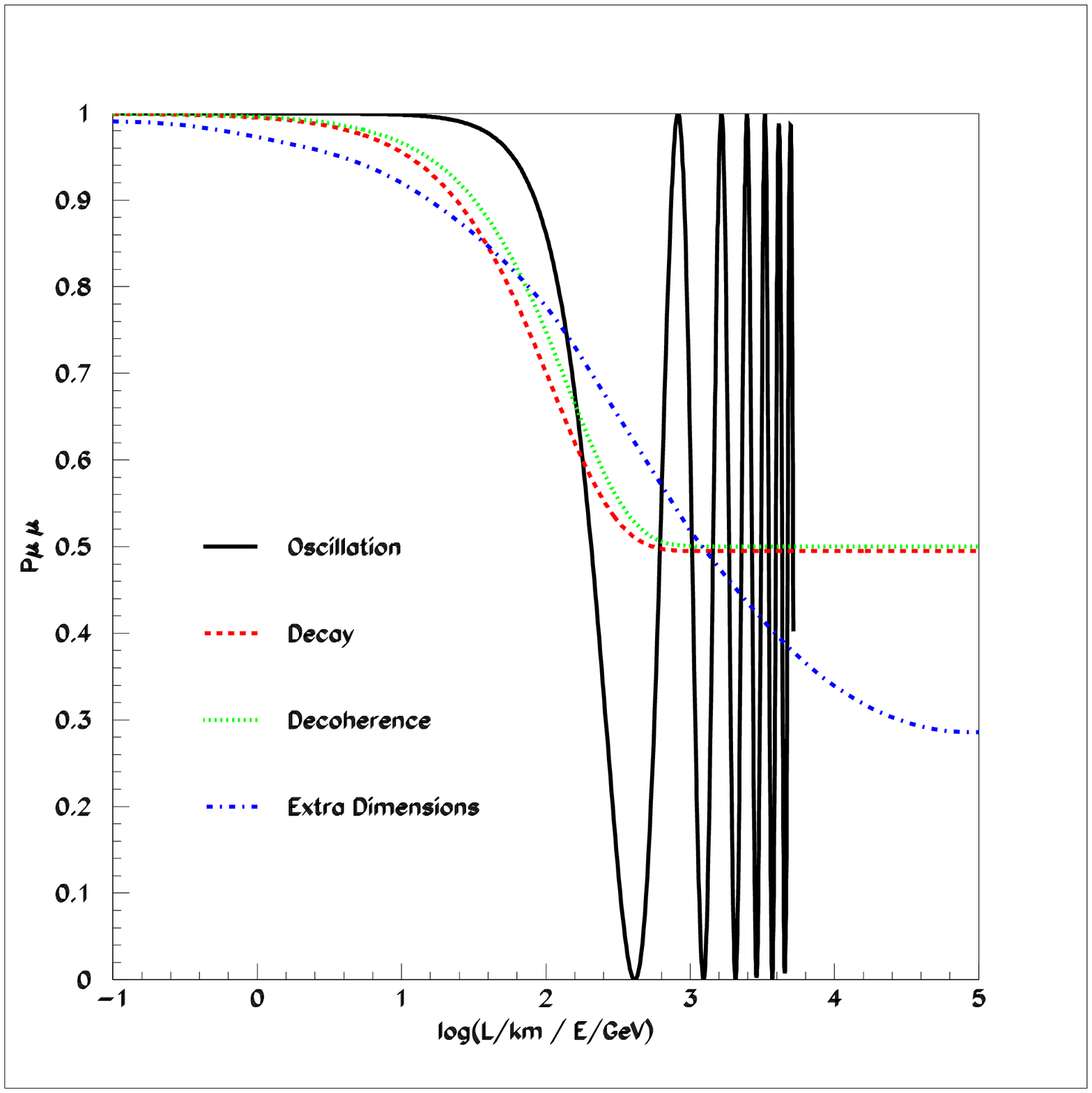}} 
\caption{Survival probality for $\nu_\mu$ versus $log_{10}$ (L/E) for
the decay model, decoherence, extra dimensions and oscillation.}
\end{figure}

\begin{figure}
\centering\leavevmode
\epsfxsize=5.5in\epsffile{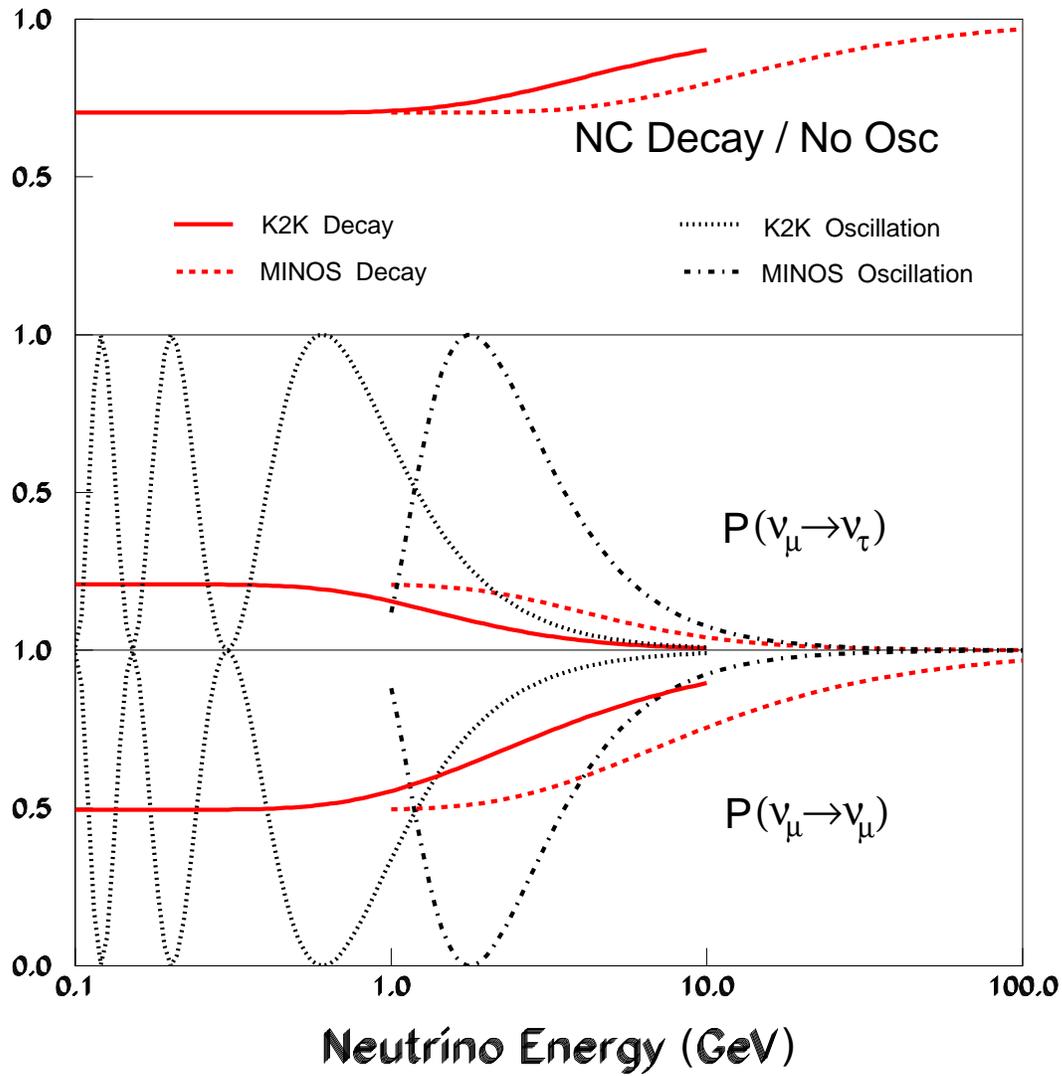}
\vspace{.5in}

\caption[]{Long-baseline expectations for the K2K and MINOS long-baseline
experiments from the
decay model and the $\nu_\mu$--$\nu_\tau$ oscillation model. The upper panel
gives the
neutral current predictions compared to no oscillations (or
$\nu_\mu$--$\nu_\tau$ oscillations).}
\end{figure}

\begin{figure}
\centering\leavevmode
\epsfxsize=5in\epsffile{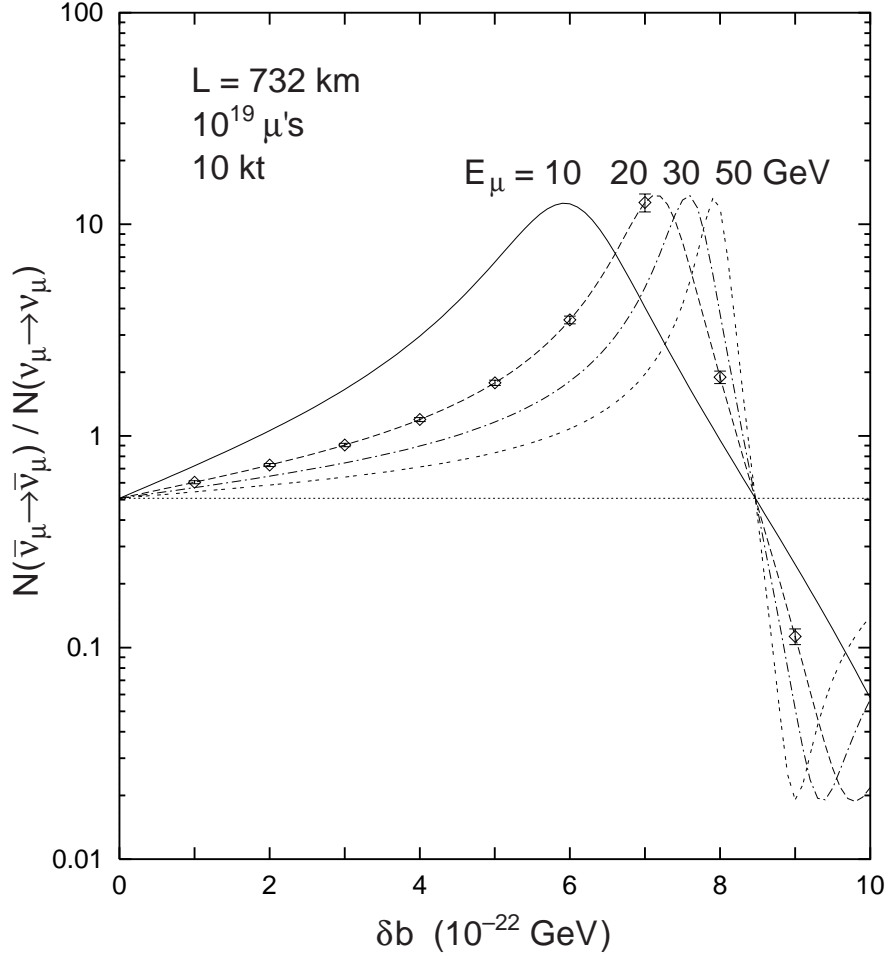}

\bigskip
\caption[]{The ratio of $\bar\nu_\mu\to\bar\nu_\mu$ to
$\nu_\mu\to\nu_\mu$ event rates in a 10~kt detector for a neutrino
factory with $10^{19}$ stored muon with energies $E_\mu = 10$, 20, 30,
50~GeV for baseline $L=732$~km versus the $CPT$-odd parameter $\delta b$
with $\theta_m = \theta_b \equiv \theta$ and phase $\eta=0$. The
neutrino mass and mixing parameters are $\delta m^2 =
3.5\times10^{-3}$~eV$^2$ and $\sin^22\theta = 1.0$. The dotted line
indicates the result for $\delta b = 0$, which is given by the ratio of
the $\bar\nu$ and $\nu$ charge-current cross sections. The error bars
are representative statistical uncertainties.}
\label{fig:ratio}
\end{figure}

\end{document}